\documentclass[12pt,a4paper]{article}
           \usepackage{times}
	   \usepackage{epsfig}
\begin{document}

  \begin{center}
	   {\Large \bf Off-lattice Kinetic Monte Carlo simulations \\[1mm]
                       of Stranski-Krastanov-like growth \/} \\
	   \ \\
	   {\sc  Michael Biehl 
	   {\rm and} Florian Much  \/} \\
	   {  Julius-Maximilians-Universit\"at W\"urzburg \/} \\
	   {   Institut f\"ur Theoretische Physik und Astrophysik \/} \\
           { and Sonderforschungsbereich 410 } \\
	   {  Am Hubland, D-97074 W\"urzburg, Germany \/} \\ 
   \end{center}
	
	 \begin{abstract}
          We investigate strained heteroepitaxial crystal growth in the framework
	  of a simplifying (1+1)-dimensional model by use of off-lattice 
          Kinetic Monte Carlo simulations. 
          Our modified Lennard-Jones system displays  the so-called
          Stranski-Krastanov growth mode: initial pseudomorphic growth 
          ends by the sudden appearance of strain induced multilayer islands 
          upon a persisting wetting layer.
          \end{abstract}

          \normalsize 
  
	  \section{Introduction}
	  In addition to its technological relevance, epitaxial 
	  crystal growth is highly attractive from a theoretical point of view. 
	  It offers many challenging open questions and provides a workshop
	  in which to develop novel methods for the modeling and simulation
	  of non-equilibrium systems, in general.

	  In particular, strained heteroepitaxial crystal growth attracts
	  significant interest as  a promising technique for the production of,
          for instance, high quality semiconductor films.
          Recent overviews of experimental and
	  theoretical investigations can be found in, e.g.,  \cite{pimpinelli,politi,liu,thisvolume}.
	  A particular attractive aspect is the 
	  possibility to exploit self--organizing phenomena for the 
	  fabrication of nanostructured surfaces  by means of 
	  Molecular Beam Epitaxy (MBE) or similar techniques.
	  
	  In many cases, adsorbate and substrate materials crystallize
	  in the same lattice but with different bulk lattice constants.
	  Frequently, one observes that the adsorbate grows {\sl layer by layer\/}, 
          initially, with the lateral spacing of atoms adapted to the substrate.
	  The misfit induces compressive or tensile strain in this {\sl pseudomorphic \/} 
	  film and, eventually, misfit dislocations will appear. These relax the strain
          and the adsorbate grows with its natural lattice constant far from
          the substrate, eventually.

	  Dislocations will clearly dominate strain relaxation
	  in sufficiently thick films and for large misfits. 
          In material combinations
	  with relatively small misfit an alternative effect governs the 
	  initial growth of thin films: Instead of growing layer by layer, the
	  adsorbate aggregates in three-dim.\ structures. The term {\sl 3D-islands\/}
	  is commonly used to indicate that these structures are spatially separated.
	  The situation is clearly different from the emergence of {\sl mounds\/} 
	  due to the Ehrlich-Schwoebel  (ES) instability \cite{pimpinelli,politi}, for instance.

	  At least two different growth scenarios display 3D-island formation:
          In {\sl Volmer-Weber\/} growth, such 
	  structures appear immediately upon the substrate when depositing the 
	  mismatched material. 
	  The situation resembles the formation of non-wetting droplets of 
	  liquid on a surface. It is frequently observed in systems where 
	  adsorbate and substrate are fundamentally different, an  example being
	  Pb on a graphite substrate \cite{pimpinelli}.

	  In the following we concentrate on the so--called {\sl Stranski-Krastanov\/}  (SK)
	  growth mode, where 3D-islands are found upon a pseudomorphic wetting-layer
	  (WL) of adsorbate material \cite{pimpinelli,politi,liu,thisvolume}.
	  Two prominent prototype SK--systems are Ge/Si and InAs/GaAs where, as 
	  in almost all cases discussed in the literature, the adsorbate
	  is under compression in the WL.  

	  In order to avoid conflicts with more detailed definitions 
	  and interpretations of the SK growth mode in the literature we will 
	  resort to the term {\sl  SK-like growth \/} in the following.
	  It summarizes the following sequence of phenomena
	  during the deposition of a few monolayers (ML) of material: 
	    \begin{enumerate}            
	    \item The layer by layer growth of a pseudomorphic 
		   adsorbate WL  up to a {\sl kinetic thickness\/} $h_{\rm WL}^{*}$.  \\[-6mm]
	    \item The sudden appearance of 3D-islands, marking the so-called 
		   2D-3D- or SK-transition \\[-6mm]
	    \item Further growth of the 3D-islands, fed by additional deposition and 
		  by incorporation of surrounding WL atoms \\[-6mm]
	    \item The observation of separated 3D-islands of similar shapes and sizes, 
		  on top of a WL with reduced {\sl stationary thickness\/} $h_{\rm WL}$. 
	    \end{enumerate} 
            Besides these basic processes a variety of phenomena  can
            play important roles in the SK-scenario,   
	    including the interdiffusion of materials
            and the segregation of compound adsorbates.
            These effects are certainly highly relevant in many cases, see
            \cite{heyn,cullis} and other contributions to this volume \cite{thisvolume}. 
	    However, SK--like growth is observed in a variety of material systems
	    which may or may not display these specific features. 
	    For instance, intermixing or segregation should be irrelevant in the somewhat
	    {\sl exotic\/} case of large organic molecules like PTCDA 
	    deposited on a metal substrate, e.g.\ Ag(111).  
	    Nevertheless, this system displays SK--like growth in
	    excellent accordance with the above  operative definition \cite{moritz}. 
	    
	    Despite the extensive investigation of SK--growth, a complete detailed 
	    theoretical picture is still lacking, apparently. 
            This concerns in particular the nature of the 2D-3D transition.
	    One problem clearly lies in the richness of the phenomenon.
	    On the other hand, the very diversity of SK--like systems gives rise to the 
	    hope that this growth scenario might be governed by a few basic universal 
	    mechanisms.  Accordingly, it should be possible to capture
	    and identify these essential features in relatively simple prototype systems. 

	    This hope motivates the investigation of simplifying models 
	    without aiming at the reproduction of material specific details.  
	    Some of the key questions in this context are: under which conditions does a
	    WL emerge and persist? How does its thickness before and after the SK-transition
	    depend on the growth conditions? Which microscopic processes trigger and control
	    the sudden  formation of 3D-islands?  How do the island size and their
	    spatial arrangement depend on the parameters of the system? 

	    Following earlier investigations of related phenomena, e.g.\
	    \cite{wolf,faux,kew,disl},  we choose a classical 
	    pair potential ansatz to represent the interactions between atoms  in our
	    model. Here, we restrict our studies to the fairly simple case of
	    a modified Lennard-Jones (LJ) system in  $1+1$ spatial dimensions, 
	    i.e.\ growth on a one-dim.\ substrate surface. 
	    As we interprete our model as a  cross-section of the physical $(2+1)$-dim.\
	    case, we still use the common term 2D-3D transition for the formation of
	    multilayer from monolayer islands.

	    We  investigate our model by means of Kinetic Monte Carlo (KMC) simulation,
	    see e.g. \cite{newman} for an introduction and overview. 
	    This concept has proven useful in the study of non-equilibrium dynamical
	    systems in general and in particular in the context of epitaxial growth,
	    see e.g. \cite{pimpinelli,politi}. 

	    Most frequently, pre-defined lattices are used  for the representation of
	    the crystal.  So-called Solid-On-Solid (SOS) models which neglect lattice defects, 
	    bulk vacancies, or dislocations have been very 
            successful in the investigation of various relevant phenomena, including
	    scenarios of kinetic roughening or mound formation due to instabilities \cite{pimpinelli,politi}.  
	    There is, however, no obvious way of including  mismatch and strain
	    effects in  a lattice gas model. A potential route is to introduce 
	    additional elastic interactions between neighboring atoms in an effective 
	    fashion. In fact, such models of hetero-systems have been studied in some of
	    the earliest  Kinetic Monte Carlo simulations of epitaxial 
	    growth \cite{madhukar,madhukar2}, see \cite{khor} for a recent example
	    of a so-called {\sl ball and spring} model.  
	    In alternative approaches the strain field of a given configuration is
	    evaluated using elasticity theory as, for instance, in \cite{meixner}.   

	    In order to account for strain effects more faithfully, including
	    potential deviations from a perfect lattice structure, it is essential  
	    to allow for continuous particle positions.
	    Given at least an approximation for the interatomic potentials, a Molecular
	    Dynamics (MD) type of simulation \cite{chemistry} would be clearly most realistic and desirable,
	    see \cite{dong} for one example in the context of heteroepitaxial growth. 
	    However, this method suffers generally from the restriction to 
	    short physical times on the order of $10^{-6}s$ or less. 
            MBE relevant time scales of seconds or minutes
	    do not seem feasible currently  even when applying sophisticated 
            acceleration techniques \cite{voter}. 
	 
	    Here, we put forward  an  {\sl off-lattice} KMC method which 
	    has been introduced in \cite{schindler,wolf,disl}
	    and apply it in the context of the SK-scenario. 
	    Some of the results have been published previously in less 
            detail \cite{skepl}.
	    The paper is organized as follows:  in the next section we  
	    outline the model and simulation method.
	    Before analysing the actual SK-like growth in section 
	    \ref{sksection} we present some basic results concerning various
            diffusion scenarios in section \ref{diffsection}. 
	    In the last section we summarize and discuss open questions and 
            potential extensions of our work.   

	   \section{Model and method} 
	   In our off-lattice model we consider pairwise interactions given by LJ-potentials
	   of the form \cite{chemistry}
	   \begin{equation} \textstyle
                   U_{ij} (U_o,\sigma) \, = \, 4 \, U_o \, \left[ \, 
		  \left( \frac{\sigma}{r_{ij}} \right)^{12}  \, - \, 
		  \left( \frac{\sigma}{r_{ij}} \right)^{6}  \, \right],
	   \label{lj}
	   \end{equation} 
	   where the relative distance $r_{ij}$  of particles $i$ and $j$ can vary continuously.  
           As a widely used approximation we cut off interactions for $r_{ij} > 3 \sigma$.

	   The choice of the parameters $\left\{U_o,\sigma\right\}$ in Eq.\ (\ref{lj}) 
           characterizes the different material properties in our model: interactions between 
	   two substrate (adsorbate) particles are specified by the sets
           $\left\{ U_s, \sigma_s\equiv 1 \right\}$ and $\left\{ U_a, \sigma_a \right\}$,
           respectively.  Instead of a third independent set we set 
	   $U_{as} = \sqrt{U_s U_a}, \sigma_{as} = \left(\sigma_s + \sigma_a\right)/2$
	   for the inter-species interaction. 

 	   As the lattice constant of a monoatomic LJ crystal is proportional to $\sigma$, 
	   the relative misfit is given by  
	   $ \epsilon \, =  \left(\sigma_a - \sigma_s\right)/{\sigma_s}$.
	   Here we  consider
	   only cases with $\sigma_a > \sigma_s$, i.e. positive misfits $\epsilon$ on the
	   order of a few percent.   
	   For the simulation of SK-like growth we set 
	   $U_s >  U_{as} >   U_{a} $.  In such systems, the formation of a WL  
	   and potential layer by layer growth should be favorable, in principle.     
           If not otherwise specified, we have set 
           the misfit to $\epsilon =4\%$, a typical value for SK-systems, and  used
           the LJ-prefactors $U_{s} = 1.0 eV, U_{a} = 0.74 eV, U_{as}\approx 0.86 eV$.
	 
	   Growth takes place on a substrate represented by six atomic layers,
	   with the bottom layer fixed and periodic horizontal boundary conditions. 
	   In the following we mostly refer to  systems with $L=800$ 
	   particles per substrate layer, additional simulations with $L=400$ or $600$ 
	   revealed no significant $L$-dependence of the results presented here. 

	   The deposition of single adsorbate particles is performed with a rate
           $R_d = L\, F$, where $F$ is the deposition flux. 
           As we interprete the substrate lattice constant as our 
	   unit of length,  flux and deposition rate (measured in ML$/s$) 
	   assume the same numerical values. 

	   The rates  of all other significant changes of the configuration  are 
	   given by Arrhenius laws.  We consider only
	   hopping diffusion events at the surface and neglect bulk diffusion, 
           exchange processes or other concerted moves.
           Furthermore,  diffusion is restricted to adsorbate particles at the 
           surface whereas jumps of substrate particles onto the surface are not considered.
	   As we will demonstrate in a forthcoming publication, these simplifications are
	   justified for small misfits $\epsilon$ in the LJ-system  because the corresponding
           rates are extremely low, see also \cite{schindler,wolf}.  

           The rate $R_i$ for a particular 
	   event $i$ is taken to be of the form  
	   \begin{equation} \label{arrhenius}  \textstyle
           R_i \, = \, \nu_o \, \exp\left[ - \frac{ E_i } { k_B \, T } \right] 
	   \end{equation} 
	   where $T$ is the simulation temperature and $k_B$ the Boltzmann constant. 
	   For simplicity, we assume that the attempt frequency $\nu_o$  is the same for all
	   diffusion events. In order to relate to physical units 
	   we use $\nu_o = 10^{12} s^{-1}$ wherever numerical  results are given. 
	   The activation barriers $E_i$  are calculated {\sl on--line\/}  given
	   the actual configuration of the system.
           This can be done by a minimal energy path saddle point calculation \cite{faux,chemistry}. 
           Here, we use a {\sl frozen crystal \/} approximation
           which speeds up the calculation of barriers significantly,    
           see \cite{schroeder} for an application in the context of 
           strained surfaces. 
           Note that the calculations are particularly simple in 1+1 dimensions: the path
           between neighboring local minima of the potential energy is uniquely determined and
           the transition state corresponds to the separating local maximum.

           An important modification concerns interlayer diffusion.
           LJ-systems in 1+1 dimensions display a strong additional  
           barrier which hinders such moves at terrace edges \cite{pimpinelli,politi}.            
           This so--called Ehrlich-Schwoebel (ES) effect is by far less pronounced  in 
           (2+1)-dim.\ systems, because interlayer moves follow a path through
           an energy saddle point rather than the pronounced maximum at the island edge.    
           In our investigation of the SK-like scenario we  remove the ES-barrier
           for all interlayer diffusion events {\sl by hand\/}.  One 
           motivation is the above mentioned
           over-estimation in one dimension.  More importantly, we wish to
           investigate strain induced island formation without interference of the ES instability.
           Note that the latter leads to the formation of mounds even in homoepitaxy 
           \cite{pimpinelli,politi}.
              
           The rates for deposition and diffusion are used 
           in a rejection-free KMC simulation \cite{newman}.   
           Using a binary search tree technique,
           one of the possible events $j$ is drawn with probability  $R_j \, / \, R$,
           where  $R= (R_d + \sum_i R_i)$ is the total rate of all potential changes.
           Time is advanced by a random  interval $\tau$ according to the
           Poisson distribution $P(\tau) = R e^{-R\tau} $
           \cite{newman}.  
            
           In order to avoid the artificial accumulation of strain due to inaccuracies
           of the method, the potential energy should be taken to the nearest 
           local minimum by variation of all particle positions in the system.
           This relaxation process affects both, adsorbate and substrate atoms. 
           In order to reduce the computational effort, we restrict the variation  to 
           particles within a radius $3 \sigma_s$ around the location 
           of the latest event, in general.  The global minimization procedure is 
           performed only after a distinct number of steps.  It is important to note, that 
           both procedures do not lead to a substantial rearrangement.  
           Significant changes of the activation energies due to relaxation
           signal the necessity to perform the global procedure more frequently.

     \section{Diffusion processes} \label{diffsection}
     Before analysing the SK-like scenario, we compare the barriers
     for hopping diffusion in various settings on the surface. 
     The investigation of systems like  Ge/Si(001) reveal  a very complicated scenario 
     due to anisotropies and the influence of
     surface reconstructions \cite{voigtlaender,voigtlaender2}. 
     For Ge on Ge(111) the 
     barrier for hopping diffusion is higher on the surface of a compressed crystal, whereas 
     diffusion is faster on relaxed Ge \cite{voigtlaender2}.

     LJ- or similar models do not reproduce this feature, in general.
     Schroeder and Wolf \cite{schroeder} consider (2+1)-dim.\ single 
     species LJ-systems and 
     evaluate the diffusion barrier for a single adatom on surfaces 
     in various lattice types.
     Among other results they find that mechanical compression of the 
     crystal lowers 
     the barrier for surface diffusion.
     However, in the mismatched two species system, it is more
     important to compare diffusion on (a) the substrate, (b) the WL 
     and (c) the surface of partially relaxed islands.

     The strong adsorbate-substrate interaction $(U_{as}>U_{a})$ 
     favors the formation of a WL,
     but it also yields deep energy minima and a relatively high 
     diffusion barrier for adsorbate 
     particles on the substrate.
     This effect is much weaker for particles on a complete wetting 
     monolayer of adsorbate and, hence, the corresponding diffusion 
     barrier is significantly lower.
     The faster diffusion further stabilizes the WL, as 
     deposited particles will reach and fill in gaps easily.  
     In principle, the trend extends to the following layers. However,
     due to the short 
     range nature of the LJ-potential the influence of the substrate
     essentially vanishes on WL of three or more monolayers.  
     With the example choice $\epsilon = 4\%$ and
     $U_{s} = 1.0 eV, U_{a} = 0.74 eV,  U_{as} \approx 0.86 eV$   
     we
     find an activation barrier of $E_a^{0}  \approx 0.57 eV $
     for adsorbate diffusion on the substrate 
     and $E_a^{1} \approx E_a^{2} \approx 0.47 eV$ 
     for diffusion on the first and second adsorbate layer, 
     respectively.  

    \begin{figure}[t]
    \begin{center}
     \includegraphics[scale=0.34]{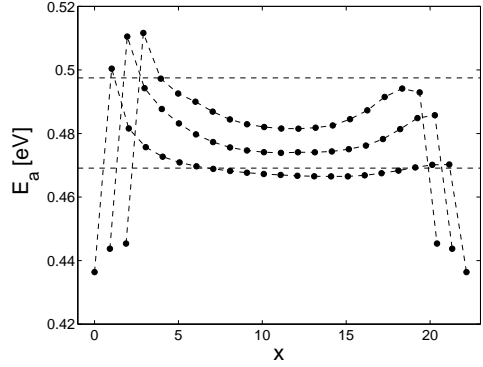}
     \caption{\label{islandbarrier}
         Diffusion barriers as obtained in our model for a single
         adatom on a flat  symmetric multilayer island
         with $24$ base particles and height $1,3,5$ layers (bottom to top curves).
         Symbols represent the activation energies for hops from the particle position
         $x$ to the left neighbor site. 
         The island is placed on top of a single WL, 
         interaction parameters are given in the text of section \ref{diffsection}.
         The leftmost barriers correspond to downward jumps at the 
         island edge with suppression of the ES-effect. 
         Horizontal lines mark the barrier for adatom diffusion 
         on the WL (lower line) and
         on perfectly relaxed adsorbate material (upper line).  
         }
    \end{center}
    \end{figure}

     For the SK-scenario the diffusion on islands of finite extension 
     is particularly
     relevant. Figure  \ref{islandbarrier} shows the barriers for 
     diffusion 
     hops on islands of various heights located upon a wetting monolayer. 
     We wish to point at two important features:
     (1) Diffusion on top of islands is, in general, slower than on the WL and
         the difference increases with the island height. In our model, this is
         an effect of the partial relaxation or over-relaxation in the island top layer. 
     (2) Depending on the lateral island size and its height,
         there is a more or less pronounced diffusion bias towards the island center,
         reflecting the spatially inhomogeneous relaxation.  
         A similar  effect has been observed in $(2+1)$ dim.\ LJ-systems \cite{schroeder}.

     Note that (2) has to be distinguished from the diffusion bias imposed by
     the ES-effect, which would be present even in 
     homoepitaxy and with particle positions restricted to a perfect undisturbed lattice. 

    \begin{figure} 
    \begin{center}
     \includegraphics[scale=0.30]{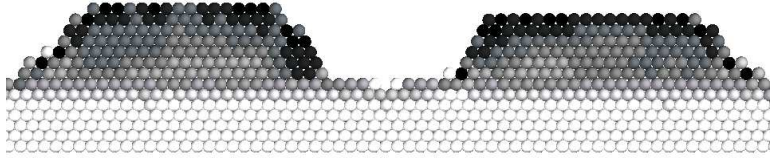}
    \caption{\label{islandpic}
         A section of a simulated crystal as obtained for  
         $R_d = 7.0 ML/s$ and $T=500K$. 
         Islands are located on a stationary WL with $h_c\approx 1$,
         the six bottom layers represent the substrate.  The darker
         a particle is displayed, the larger is the average distance
         from its nearest neighbors. } 
    \end{center}
    \end{figure}

    Clearly, (1) and (2) favor the formation of islands upon islands and hence 
    play an important role in SK-like growth. They concern adatoms 
    which are deposited directly onto the islands as well as particles that hop
    upward at edges, potentially. 
    As we will argue in the following section, upward diffusion moves play
    the more important role in the 2D-3D transition of our model.

   \section{SK-like growth scenario}  \label{sksection}
   In our investigation we follow a scenario which is frequently studied
   in experiments \cite{pond,seifert}:  In each simulation run a total of 4 ML adsorbate 
   material is deposited at rates in the range
   $ 0.5 ML/s \leq R_d \leq 9.0 ML/s$. 
   After deposition is complete, a 
   relaxation period with $R_d=0$ of about $10^7$ diffusion steps follows, corresponding to
   a physical time on the order of $0.3s$.   

   As we have demonstrated in \cite{disl}, strain relaxation through dislocations  
   is not expected for, say,  $\epsilon = 4\%$ within the first few adsorbate layers.  
   Indeed misfit dislocations were
   observed in none of the simulations presented here. 
   Results have been obtained on average over at least 15 independent simulation runs
   for each data point. 
   
   In our simulations we observe the complete scenario of SK-like growth
   as described in  the introduction.  Illustrating mpeg movies of the simulations 
   are available upon request or directly at our web pages \cite{web}. 
   A section of a simulated crystal after deposition and island formation
   is shown in Figure \ref{islandpic}.   
   Ultimately, the formation of islands is driven by the relaxation of strain. 
   As shown in the figure, material within
   the 3D-island and at its surface can assume a lattice constant close to
   that of bulk adsorbate. On the contrary, particles in the WL are {\sl forced\/}
   to adapt the substrate structure.

\begin{figure} 
    \begin{center}
     \includegraphics[scale=0.34]{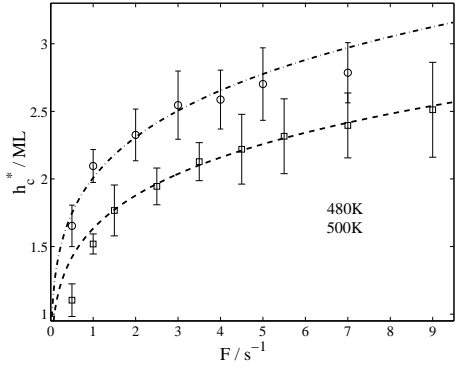} 
    \caption{\label{hcfig}
    Kinetic critical WL thickness $h_{\rm WL}^*$ {\sl vs.\/} the deposition flux 
    for two different substrate temperatures. Both curves correspond to Eq.\ (\ref{gamma})
    with $h_o$ and $\gamma$ obtained from the data of $T=500K$, only.  } 
    \end{center}
    \end{figure}

   During deposition, monolayer islands located on the WL undergo a
   rapid transition to bilayer islands at a well-defined thickness $h^*_{\rm WL}$. 
   For the systematic determination of $h^*_{\rm WL}$ we follow \cite{pond} 
   and fit the density $\rho$ of 3D-islands as $\rho = \rho_o \, (h-h_{\rm WL}^*)^\alpha$, 
   finding comparable values of the exponent $\alpha$. 
   Figure \ref{hcfig} displays the results for two different substrate temperatures $T$
   and various deposition rates $R_d$.
   The increase of $h^*_{\rm WL}$ with decreasing $T$  agrees qualitatively  
   with several experimental findings, see \cite{seifert} as one example.
   Heyn discusses the  effect of adsorbate/substrate interdiffusion  
   on the $T$-dependence of $h^*_{\rm WL}$ in the InGaAs system \cite{heyn}. 
   Reassuringly, our result is consistent with his findings for absent 
   intermixing. 

   A key observation is the significant increase of $h_{\rm WL}^*$ with increasing 
   deposition flux. Qualitatively the same flux dependence is reported for InP/GaAs  
   heteroepitaxial growth in \cite{seifert}. 
   This behavior leads to the conclusion that the emergence of islands upon
   islands, i.e. the SK-transition, is mainly due to particles performing 
   upward hops onto existing monolayer islands.
   If, on the contrary, the formation of second or third layer nuclei by  
   freshly deposited adatoms was the dominant process, one would expect more frequent
   nucleation  and an earlier 2D-3D transition at higher growth rates. 
   Note again that we have suppressed the ES-effect explicitly which would also lead
   to more frequent mound formation at higher deposition rates \cite{pimpinelli,politi}.

    \begin{figure} 
    \begin{center}
     \includegraphics[scale=0.34]{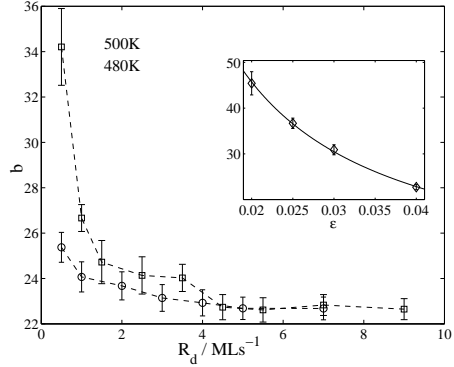}
    \caption{ \label{sizefig}
     Average base size $\left\langle b \right\rangle$   
     of multilayer islands as a function of $R_d$ at $T=500K$, together
     with standard error bars. The inset shows the result for $R_d= 4.5 ML/s$  
     and different misfit parameters, the solid line corresponds to 
     $\left\langle b \right\rangle = 0.91/\epsilon$.  } 
    \end{center}
    \end{figure}

    In our model, upward diffusion is the limiting effect and 
    sets the characteristic time for the SK-transition.  The frequency of these 
    processes strongly depends on the temperature and it has
    to be compared with the deposition rate. A high incoming flux  will fill layers
    before particles can perform upward hops, and hence it will delay the 
    2D-3D-transition, cf. \ref{hcfig}.
    These considerations, together with the arguments  of \cite{seifert}, suggest
    a functional dependence of $h_{\rm WL}^*$ on $F$ and $T$  of the form
    \begin{equation} \label{gamma} \textstyle
        h_{\rm WL}^* \, = \, h_o \, \left( \frac{F}{R_{up}} \right)^\gamma,  
    \end{equation}
    where $R_{up}$ is the Arrhenius rate for upward hops. 
    In our simulations we find a typical value which corresponds to an 
    activation barrier $E_{up} \approx 1.0 eV$ close to the transition.
    Using this value we have performed a non-linear fit according to Eq.\ 
    (\ref{gamma}) based on the data for $T=500 K$. Inserting the obtained 
    parameters $h_o \approx 3.0 ML$ and $\gamma \approx 0.2$ in
    Eq.\ (\ref{gamma}) yields good agreement for $T=480 K$, as well.  
    Note that the value of $\gamma$ is expected to vary 
    with the material system in the range
    $0<\gamma<0.5$ according to \cite{seifert}.  
    
    Our assumption that upward diffusion is more important than 
    nucleation of deposited adatoms on top of islands 
    is further supported by the observation that multilayer islands tend to form
    on a  WL of, say, two monolayers thickness even without particle deposition.
    
    After the 2D-3D-transition,  islands grow by incorporating newly deposited 
    material, but also by consumption of the surrounding WL. 
    Note that very large islands are observed to split by means of 
    upward diffusion events onto their top layer \cite{web}.
    The migration of WL particles towards and onto the islands can also
    extend into the relaxation period after deposition ends. 
    Eventually, a stationary WL thickness $h_{\rm WL} \approx 1$ is observed
    in our example scenario with $U_{as} \approx 0.86 eV$.  By increasing the 
    strength of the adsorbate/substrate interaction  we can achieve, e.g.,
    $h_{\rm WL} \approx 2$ for $U_{as} \approx 2.7 eV$,
    but it is difficult to stabilize a greater stationary
    thickness, due to the short range nature of the LJ-potential.  

    Finally, we discuss the properties of 3D-islands emerging in SK-growth.    
    As an example, their base length $b$ is measured as the number of particles 
    in the island bottom layer. 
    The results discussed in the following were obtained  at the end
    of the relaxation period with $R_d=0$. Whereas mean values do not
    change significantly, fluctuations are observed to decrease with time  
    in this phase. 

    We observe that, for fixed temperature,
    the island size decreases with increasing deposition flux,
     as  observed in several experimental studies 
    \cite{seifert}. 
    However, $\left\langle b \right\rangle$ 
    becomes constant and independent
    of $T$ for large enough deposition rate, cf. Figure \ref{sizefig}.    
    A corresponding behavior is found for the island density and their
    lateral distance, hinting at a considerable degree of spatial ordering \cite{skepl}.
    The saturation behavior further demonstrates the importance 
    of upward hops {\sl vs.\/} aggregation of deposited particles on islands.
    The latter process would yield a continuous increase of the island
    density with $R_d$.

    We find a narrower distribution of island sizes with  
    increasing $R_d$  \cite{skepl}.
    The island size distribution in the saturation
    regime  will be studied 
    in greater detail in forthcoming investigations.

    We conclude our discussion by noting that, in the saturation regime
    of high growth rates, the typical island size follows a simple power
    law: $\left\langle b \right\rangle \propto  1/\epsilon$, cf.  Figure \ref{sizefig}.  
    Very far from equilibrium, the only relevant length scale in the system
    is given by the relative periodicity $1/\epsilon$ of the adsorbate and
    substrate lattices. This characteristic length was already found to
    dominate in the formation of misfit dislocations for larger values of $\epsilon$
    \cite{disl}.

   \section{Conclusion and Outlook}

    Despite its conceptual simplicity  and the small number of free
    parameters, our model reproduces various phenomena of 
    heteroepitaxial growth.
    We believe that, with a proper choice of interaction parameters, our model
    should be capable of reproducing all three prototype growth modes:
    layer by layer growth (for very small misfits), Volmer-Weber (for
    $U_{as} < U_{a}$) and, as demonstrated here, the Stranski-Krastanov mode. 
    Our work  provides a fairly detailed and plausible picture of the latter. The
    key features are:  \\
    (a)  The strong adsorbate/substrate interaction favors the
         WL formation  and results in a relatively slow diffusion
         of adatoms on the substrate. Diffusion on the WL is significantly
         faster and the corresponding barrier decreases with the WL thickness. \\
    (b)  Strain relaxation leads to a  pronounced bias towards
         the island center on top of finite mono- and multilayer islands located on
         the WL.  In addition, diffusion is slower on top of the partially relaxed islands
         than on the WL, in our model.  

     Whereas (a) favors the formation and persistence of the WL, 
     (b) clearly de-stabilizes layer by layer growth. 
     We find that the microscopic process which 
     triggers the transition is upward diffusion of adatoms from the 
     WL and at island edges.
     The corresponding barriers  decrease with the WL thickness analogous to 
     (a).  As a result of the competing effects, the 2D-3D-transition occurs at a  critical
     thickness which depends upon $T$ and $R_d$ as suggested in  Eq.\ (\ref{gamma}). 

     We hypothesize  that strain effects induce spatial
     modulations with a characteristic length scale $\epsilon^{-1}$ and thus 
     control the island size far from equilibrium.  
     We find a corresponding saturation regime for large enough deposition fluxes. 
     The precise mechanism of the size selection will be the subject 
     of a forthcoming project.  
     A related open question concerns the crossover from 
     kinetically controlled island sizes to the equilibrium behavior
     which should be achieved in the limit of very small $R_d$.
     Several arguments \cite{politi,meixner} suggest that
     the typical island size close to equilibrium should be of order $\epsilon^{-2}$.  
     
     Further investigations will concern the effects of
     intermixing and segregation which have been excluded from our 
     model, so far.  To this end we will consider  the
     co-deposition of both species  and  allow for exchange diffusion
     at the substrate/adsorbate interface.  
     
     In order to test the potential universality of our results, we
     will introduce different types of interaction potentials in our
     model.  Ultimately, we plan to extend our model to the  relevant  
     case of $2+1$ dimensions and to more realistic empirical potentials
     for semiconductor materials, e.g.\ \cite{tersoff}.

     \paragraph{Acknowledgment:} 
     F. Much has been supported by the Deutsche Forschungsgemeinschaft.
     M. Biehl thanks the organizers and all participants of the
    {\sl NATO ARW on Quantum Dots: Fundamentals, Applications, and Frontiers\/} \cite{thisvolume}
    for pleasant and useful discussions. 
     We also thank B. Voigtl\"ander for communicating the results of
     \cite{voigtlaender2} prior to publication.

    \end{document}